# AN EFFICIENT METHOD BASED ON GENETIC ALGORITHMS TO SOLVE SENSOR NETWORK OPTIMIZATION PROBLEM

Ehsan Heidari[1] and Ali Movaghar[2]

[1]Department of Computer Engineering, Islamic Azad University, Doroud, Iran
`Heidari@iau-doroud.ac.ir`

[2]Department of Computer Engineering, Sharif University of Technology, Tehran, Iran
`Movaghar@sharif.edu`

*ABSTRACT*

*Minimization of the number of cluster heads in a wireless sensor network is a very important problem to reduce channel contention and to improve the efficiency of the algorithm when executed at the level of cluster-heads. In this paper, we propose an efficient method based on genetic algorithms (GAs) to solve a sensor network optimization problem. Long communication distances between sensors and a sink in a sensor network can greatly drain the energy of sensors and reduce the lifetime of a network. By clustering a sensor network into a number of independent clusters using a GA, we can greatly minimize the total communication distance, thus prolonging the network lifetime. Simulation results show that our algorithm can quickly find a good solution.*

## 1. INTRODUCTION

Recent advances in micro-electro-mechanical systems (MEMS) technology, wireless communications, and digital electronics increased the development of low-cost, low-power and multifunctional sensor nodes that are small in size and communicate in short distances [1], [16], [10]. These tiny sensor nodes, which consist of sensing, data processing, and communicating components, leverage the idea of sensor networks based on collaborative effort of large number of nodes. A wireless sensor network usually has a base station that can have a radio relation with other sensor nodes in a network. The data of each sensor node will send to the base station directly or by the other nodes. Then, all the information will gather and they will be edited in a base station for parameters like: temperature, pressure and humidity. After that we can estimate the real amount of that parameter, clearly.

A sensor network is composed of a large number of sensor nodes, which are densely deployed either inside the phenomenon or very close to it. The position of sensor nodes does not need to be engineered or pre-determined. On the other hand, it also means that sensor network protocols and algorithms must possess self-organizing capabilities. Sensor networks may consist of many different types of sensors such as seismic, low sampling rate magnetic, thermal, visual, infrared, and acoustic and radar. Sensor nodes can be used for continuous sensing, event detection, event ID, location sensing, and local control of actuators. The concept of micro-sensing and wireless connection of these nodes promises many new application areas. Sensor networks are also candidates for solving a variety of other social, military, and environmental problems, including secretly monitoring enemy activities on a military battlefield located in an inhospitable terrain; and detecting wildfires in a densely forested area[21] .

In designing the wireless sensor networks there are some limitations like; the small size, low weight, use of energy and the low prices of sensors [4]. Among these factors the amount and manner of using energy is very important, too. On the other hand the decreased use of energy in wireless sensor networks has a direct relation with their increased longevity and this issue is very important itself. For increasing the longevity of a network the load of energy is distributed equally between the nodes of a network. On the other hand the load distribution is monotonous. The monotonous distribution of energy load in a network will cause the network nodes lose their energy in a short distance of each other, and the chronological distance between the dying of first node and the last node will become shorter. The monotonous distribution of energy load and increasing the longevity of a network comprise this point that the sending data of a network is collected from the environment monotonously.

The relational protocols have an important role in efficiency and longevity of wireless sensor networks [2]. So, designing efficient protocols for using energy is the necessity for wireless sensor networks. By using these protocols not only the whole consumed energy in a network will decrease, but also the load of consumed energy will be distributed among the network nodes monotonously. So the longevity of the network will increase. Among the available protocols, Hierarchical protocols are greatly economized in consuming energy by the network [17]. In these protocols the network will be divided into several clusters and in each of these clusters one node will be introduced as a head cluster. The tasks of these head clusters are gathering the sending data from the nodes of the cluster, omitting the repetitious data, mixing the data and sending these data to sink. In these protocols selecting a node as a cluster head and mixing the data are greatly efficient in increasing the scalability and longevity of a network. Up to now, several clustering protocols like: LEACH [7], [8], TEEN [14], APTEEN [13], DBS [3], EMPAC [12], FTPASC [11] and sop [20] have been given for the wireless sensor networks.
In this article, we suppose the sensor network like static. The sensors are distributed monotonously in an environment and they are far a way from sink. The place of sensors can be measured by the GPS system.

Clustering a network for keeping the total distance to a minimum is the NP-HARD issue. One genetic algorithms is the algorithm for efficient searching that imitates a transforming procedure from environment. Of course, the genetic algorithm can be used completely in many NP-hard problems like Optimization and TSP (The Traveling Salesman Problem).
Here, we suggest a GA for clustering sensor nodes and selection of the minimum of clusters. In this way the total of connecting distances and the use of energy will decrease effectively and the longevity of network will increase.

The rest of the article is as follows. Some methods of previous clustering are presented in chapter 2, briefly. In chapter 3 some primary concepts about genetic algorithm are introduced. In Chapter 4 the suggested method is put forward. In chapter 5 the results of simulation are discussed, and finally in chapter 6 there will be a conclusion.

## 2. RELATED WORKS

Hein Zelman et al. [7], [8] describe the LEACH protocol, a hierarchical and self-organized cluster-based approach for monitoring applications. The data collection area is randomly divided into several clusters, where the number of clusters is pre-determined. Based on time division multiple accesses (TDMA), the sensor nodes transmit data to the CHs, which aggregate and transmit the data to the BS. A new set of CHs are chosen after specific time intervals. A node can be re-elected only after all the remaining candidates have been elected. Lindsey et al. [19] proposes PEGASIS, an extension of LEACH, in which nodes transmit to their nearest neighbors and eventually transmit the messages to the base station. Bandyopadhyay and Coyle [18] describe a multi-level hierarchical clustering algorithm, in which the parameters for minimum energy consumption are obtained using stochastic geometry. Ye et al. [22] describes a contention-based and medium access protocol, SMAC, which reduces energy consumption by using virtual clusters. Common sleep schedules are developed for the clusters, and in-channel signaling is used to avoid collisions. Tillett et al. [9] proposes the PSO (Particle Swarm Optimization) approach to divide a sensor node field into groups of nodes which are equal in size. PSO is an evolutionary programming technique that mimics the interaction of ants or termites to find a good solution. Although dividing into equal sized clusters balances the energy consumption of cluster heads, this method is not applicable to some networks where nodes are not evenly distributed.

All of the above algorithms are assumed a fixed number of clusters. When we do not know the number of clusters in advance; we would like to solve much harder problems. Our approach uses a GA to determine both the number and location of the cluster heads. This approach minimizes the communication distance in a sensor network; too.

## 3. BASIC CONCEPTS

### 3.1. Genetic Algorithms [6], [15]

Genetic algorithms (GAs) are adaptive methods which may be used to solve search and optimization problems. They are based on the genetic processes of biological organisms. According to the principles of natural selection and survival of the fittest; natural populations are evolved in many generations. By mimicking this process, genetic algorithms are able to use solutions to real world problems, if they have been suitably encoded. GAs and the population of "individuals", represent a possible solution to a given problem. Each individual assigns a "fitness score" according to how good a solution to the problem is. The highly-fit individuals are given opportunities to "reproduce", by "cross breeding" with other individuals in the population. It produces new individuals as "offspring", which share some features taken from each "parent". The least fit members of the population are less likely to be selected for reproduction, so they will "die out". A new population of possible solutions is produced by selecting the best individuals from the current "generation", and by mating them we can produce a new set of individuals. This new generation contains a higher proportion of the characteristics possessed by the good members of the previous generation. In this way, in many generations, good characteristics are spread throughout the population. By favoring the mating of the fit individuals, the most promising areas of the search space are explored. If the GA has been designed well, the population will converge to an optimal solution to the problem.

### 3.2. Basics of Genetic Algorithms [23]

The most common type of genetic algorithm works like this: a population is created with a group of individuals created randomly. The individuals in the population are then evaluated. The evaluation function is provided by the programmer and gives the individuals a score based on how well they perform at the given task. Two individuals are then selected based on their fitness, the higher the fitness, and the higher chance of being selected. These individuals then will reproduce to create one or more offspring. After that the offspring are mutated randomly.

This process continues until a suitable solution has been found or a certain number of generations have passed, depending on the needs of the programmer.

**Selection**

The selection process determines which of the chromosomes from the current population will mate (crossover) to create new chromosomes. These new chromosomes join to the existing population. This combined population will be the basis for the next selection. The individuals (chromosomes) with better fitness values have better chances of selection. While there are many different types of selection, I will cover the most common type - roulette wheel selection. In roulette wheel selection, individuals are given a probability of being selected that is directly proportionate to their fitness. Two individuals are then chosen randomly based on these probabilities and produce offspring. Pseudo-code for a roulette wheel selection algorithm is shown below (Figure 1).

```
For all members of population
    Sum += fitness of this individual
End for

For all members of population
    Probability = sum of probabilities + (fitness / sum)
    Sum of probabilities += probability
End for

Loop until new population is full
    Do this twice
        Number = Random between 0 and 1
        For all members of population
            If number > probability but less than next probability
                Then you have been selected
        End for
    End
    Create offspring
End loop
```

Figure 1. Pseudo code of the roulette wheel selection algorithm

**Crossover**

Now; you have selected your individuals, and somehow you know that you are supposed to produce offspring with them, but how you can do it? The most common solution is something called crossover, and while there are many different kinds of crossover, the most common type is single point crossover. In single point crossover, you choose a locus at which you swap the remaining alleles from one parent to the other. This is complex and is best understood visually (figure 2).

As you can see, the children take one section of the chromosome from each parent. The point at which the chromosome is broken depends on the randomly selected crossover point. This particular method is called single point crossover because only one crossover point exists. Sometimes only child 1 or child 2 is created, but oftentimes both offspring are created and put into the new population. Crossover does not always occur, however. Sometimes, based on a set probability, no crossover occurs and the parents are copied directly to the new population. The probability of crossover occurring is usually 60% to 70%.

|        | 1     |             | 2     |            |
|--------|-------|-------------|-------|------------|
| Parent | 11001 | 01100000110 | 10010 | 10110001110 |
| Child  | 11001 | 10110001110 | 10010 | 01100000110 |

Figure 2. A single-point crossover

**Mutation**

After selection and crossover, you now have a new population full of individuals. Some are directly copied, and others are produced by crossover. In order to ensure that the individuals are not all exactly the same, you are allowed for a small chance of mutation. You loop through all the alleles of all the individuals, and if that allele is selected for mutation, you can either change it by a small amount or replace it with a new value. The probability of mutation is usually between 1 and 2 tenths of a percent. A visual for mutation is shown below (figure 3).

|          | Offspring          |
|----------|--------------------|
| Original | 1100110**1**10001110 |
| Mutated  | 1100110**0**10001110 |

Figure 3. An example of mutation

As you can easily see, mutation is fairly simple. Change the selected alleles based on what you feel is necessary and then move on. Mutation is, however, vital to ensuring genetic diversity within the population.

**Fitness function**

A fitness function must be devised for each problem to be solved. Given a particular chromosome, the fitness function returns a single numerical "fitness," or "figure of merit," which is supposed to be proportional to the "utility" or "ability" of the individual which that chromosome represents. For many problems, optimization and the fitness function should simply measure the value of the function.

## 4. OUR GENETIC ALGORITHM

The BS uses a GA to create energy-efficient clusters for a given number of transmissions. The node is represented as a bit of a chromosome. The head and member nodes are represented as 1s and 0s, respectively. A population consists of several chromosomes. The best chromosome is used to generate the next population. Based on the survival fitness, the population transforms into the future generation. Initially, each fitness parameter is assigned an arbitrary weight; however, after every generation, the fittest chromosome is evaluated and the weights for each fitness parameter are updated accordingly. The GA outcome identifies suitable clusters for the network. The BS broadcasts the complete network details to the sensor nodes. These broadcast messages include: the query execution plan, the number of cluster heads, the members associated with each cluster head, and the number of transmissions for this configuration. All the sensor nodes receive the packets broadcasted by the BS and clusters are created accordingly; thus the cluster formation phase will be completed. This is followed by the data transfer phase.

## 4.1. Problem Representation

Finding appropriate cluster heads is critically important for minimizing the distance. We use binary representation in which each bit corresponds to one sensor or node. "1"means that corresponding sensor is a cluster-head; otherwise, it is a regular node. The initial population consists of randomly generated individuals. GA is used to select cluster-heads.

## 4.2. Crossover and Mutation used in this article

**Crossover**

In this article, we use one-point crossover. If a regular node becomes a cluster-head after crossover, all other regular nodes should check if they are nearer to this new cluster-head. If so, they switch their membership to this new head. This new head is detached from its previous head. If a cluster-head becomes a regular node, all of its members must find new cluster-heads. Every node is either a cluster-head or a member of a cluster-head in the network.

**Mutation**

The mutation operator is applied to each bit of an individual with a probability of mutation rate. When applied, a bit whose value is 0 is mutated into 1 and vice versa.

## 4.3. Selection

The selection process determines which of the chromosomes from the current population will mate (crossover) to create new chromosomes. These new chromosomes join to the existing population. This combined population will be the basis for the next selection. The individuals (chromosomes) with better fitness values have better chances of selection. There are several selection methods, such as: "Roulette-Wheel" selection, "Rank" selection,"Steady state" selection and "Tournament" selection.

In Roulette-Wheel, which is used in this article, chromosomes with higher fitness compared to others and they will be selected for making new offspring. Then, among these selected chromosomes, the ones with lesser fitness than others will be removed and new offspring would be replaced with the former ones.

## 4.4. Fitness Parameters

The total transmission distance is the main factor we need to minimize. In addition, the number of cluster heads can factor into the function. Given the same distance, fewer cluster heads result in greater energy efficiency.

**The total transmission distance (Total Distance) to Sink:** The total transmission distance, TD, to sink is the sum of all distances from sensor nodes to the Sink (figure 4)

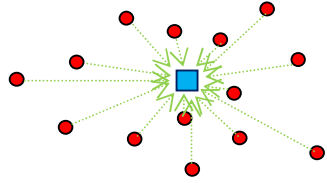

Figure 4. The all distances from sensor nodes to the Sink

**Cluster Distance (Regular nodes to Cluster head, Cluster head to sink distance):** The cluster distance, RCSD, is the sum of the distances from the nodes to the cluster head and the distance from the head to the Sink (figure 5).

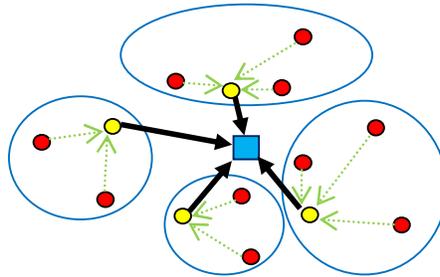

Figure 5. The sum of the distances from the nodes to the cluster head and the distance from the head to the sink

**Transfer Energy (E):** Transfer energy, E, represents the energy consumed to transfer the aggregated message from the cluster to the Sink. For a cluster with m member nodes, cluster transfer energy is defined as follows:

$$E = \sum_{i=1}^{m} TE_{CHi} + ((m-1)*ER) + TE_{CHs} \qquad (1)$$

The first term of Equation 1(TE) shows the energy consumed to transmit messages from m member nodes to the cluster head. The second term ((m-1)*ER) shows the energy consumed by the cluster head to receive m messages from the member nodes. Finally, the third term (TE) represents the energy needed to transmit from the cluster head to the sink.

**We represent the number of cluster-heads with TCH and the total number of nodes with N.**

### 4.5. Fitness Function

The used energy for conveying the message from cluster to sink and the sending distance are the main factors that we need to minimum them. In addition to these, we can insert the decreasing number of clusters in our function that it can affect the energy function like the decreased sending distance, because the clusters use more energy in spite of other nodes. So the chromosome fitness or F is a function (Fitness Function) of all the above fitness parameters that we define it like Equation (2).

$$F = \frac{1}{E} + (TD - RCSD) + (N - TCH) \quad (2)$$

As we explained here E means the essential energy for sending information from cluster to sink, TD is the total distance of all the nodes to sink, RCSD is the total distance of Regular nodes to clusters and the total distances of all the clusters to sink, N and TCH are the number of all the nodes and clusters.

For calculating this function, the amounts of N and TD are fixed, but the amounts of E, TCH and RCSD are changing.

The fewer used energy, the shorter sending distance or the fewer number of clusters will lead to the more amount of fitness value. Our suggested genetic algorithm tries to find a suitable solution by increasing the fitness value.

In relation (2), we can define the role of every parameter for the formation of clusters. For example if we use 0/E instead of 1/E it means that we do not put the energy parameter in clustering, i.e. the clusters are formed based on the distance and the number of clusters. In other mode if we use 100/E in our relation, it shows that the role of energy decrease in clustering is more noticed.

In fact, TD is the distance of all nodes to the sink, and RCSD is the total distance of nodes to the cluster head and cluster heads to the sink. If the difference amount of TD- RCSD is big, it means that we can decrease the relational distance of nodes to sink by clustering. The results will be the energy decrease for the relation of nodes and sink. For normalizing the relation, we can divide (TD-RCSD) on TD which is an always fixed. We can use (TD-RCSD)/TD instead of (TD-RCSD). When the relation becomes normal, we can multiply in a fixed number. It can lead to the importance of the distance parameter in relation.

The last part of relation leads to the importance of the numbers of cluster heads. As it was mentioned before, the nodes of cluster heads use more energy in comparison with the other nodes. Then, we must decrease the number of cluster heads to the very least.

So, we decrease the number of cluster heads from the total numbers of the nodes in each generation. If this amount of difference is increased, the number of cluster heads will decrease. For normalizing the relation, we divide this part on fixed N, I.e. we write the third part of relation, like (N-TCH)/N. If we want to increase the effects of third part or the number of cluster heads in relation calculation, we can multiply it in a fixed number.

At last, the relation (3) is suggested for calculation based on the simulations and the comparison between the effects of different values of parameters in the relation:

$$F = \frac{100}{E} + \frac{(TD - RCSD)}{TD} + \frac{(N - TCH)}{N} \quad (3)$$

It shows that in this relation, we paid more attention to the decrease of energy.

## 5. GENETIC ALGORITHM EXPERIMENTS

In this chapter, the performance of the suggested method will be evaluated. For this, we use MATLAB software. First, we compare effect of the importance of different parameters in a function. Then, we compare clustering in LEACH algorithm with a suggested method. In our experiment we suppose that the Base station with one interval is out of wireless sensor network area. The number of sensor nodes for this experiment is about 200. The used parameters in this experiment are shown in table 1.

Table 1. The used parameters in this experiment.

| PARAMETER | VALUE |
|---|---|
| N | 200 |
| Population size | 100 |
| Selection type | Roulette-Wheel |
| Crossover type | One point |
| Crossover rate | 0.8 |
| Mutation rate | 0.3 |
| Number of iteration | 200 |

The nodes are distributed in an environment like figure 6. The place of sink is considered at the middle of this environment, i.e. [0, 0].

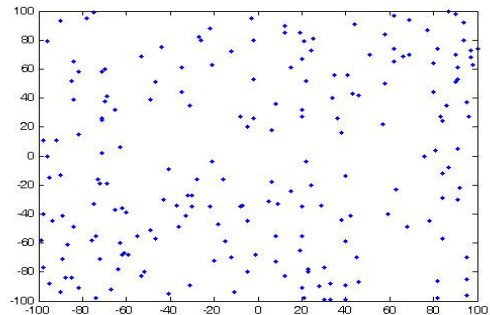

Figure 6. Distributed nodes

### 5.1. Simulation by the use of Relation (2)

First, we do the simulation by the use of relation (2) and by the help of the information of table (1).Figure 7 shows the amount of fitness function after each generation. As we can see, after each generation the suggested algorithm increases the amount of fitness and almost after 100 circuits of doing algorithm, the amount of fitness will reach to its maximum.

26

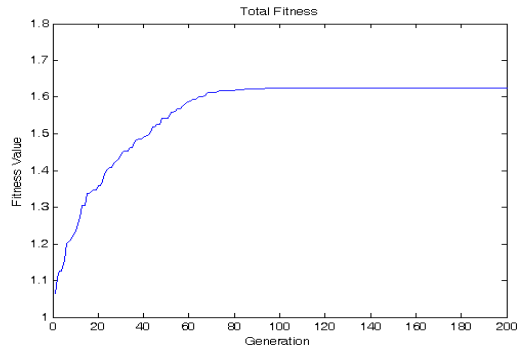

Figure 7. The amount of fitness by the use of relation (2)

The above figure shows that relation (2) can increase the amount of fitness function after each generation. The next figure is related do the number of cluster heads. As it is expected, the number of cluster heads is decreased after each generation.
As it is showed in figure (8), at first the number of cluster heads is 78, while after 140 generation the number of cluster heads is decreased to 25 and after that this number will be fixed. So, the suggested function is effective for decreasing the number of cluster heads.

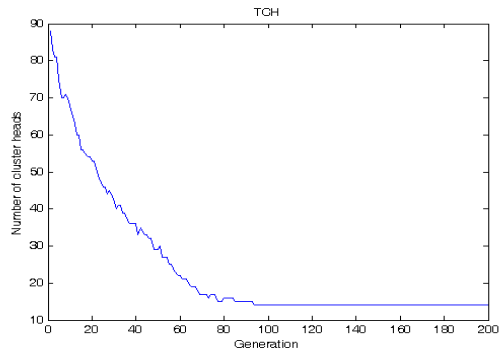

Figure 8. The number of cluster heads by the help of relation (2)

Figure 8 shows the RCSD changes. With figure 9 we can understand that the total distance is decreased after each generation.

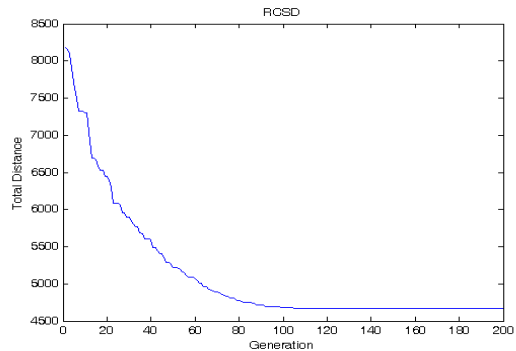

Figure9. The minimum distance by the help of relation (2)

With the figures 8 and 9 which show the decrease of the numbers of cluster heads and communication distance, we can conclude that after each generation the consuming energy must be decreased. Based on the result of the simulation we can reach to the figure 10, this figure confirms the above materials. The following figure shows the decrease of consuming energy after each generation.

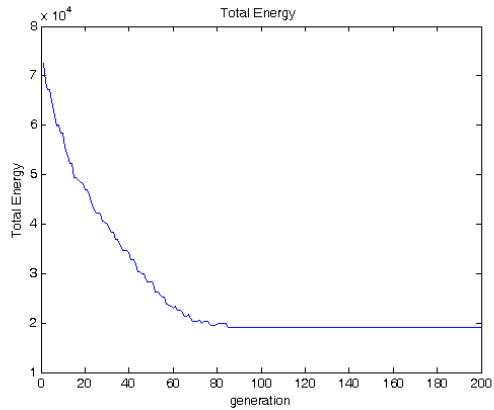

Figure 10. The whole consumed energy in a network after each iteration by the use of relation (2)

Figure 11 shows the method of cluster formation after doing the genetic algorithm and 200 times of performing relation (2). In this figure, the sink is shown at the middle of the environment with a black square. The regular nodes are shown with blue circles and cluster heads are shown by the yellow circles wich are bigger than the normal nodes.

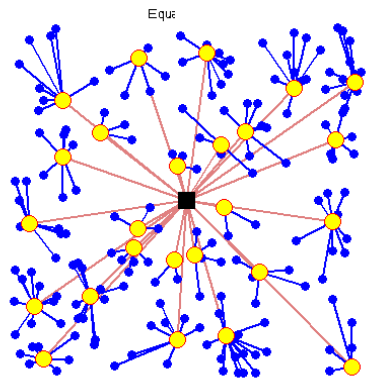

Figure 11. The formation of clusters after performing relation (3)

## 5.2. Simulation based on Relation (3)

Now, we repeat the above experiments by the use of relation (3). In this part, we extract the parameters from table (1) for simulation. Figure 12 shows the amount of fitness function after each generation. As we can see, after each generation the suggested algorithm increases the amount of fitness.

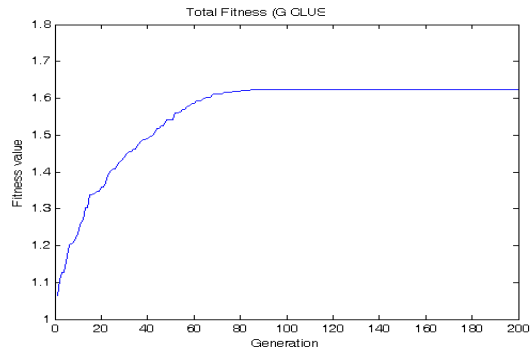

Figure 12. The amount of fitness by the help of relation (3)

Cluster heads use more energy, so by decreasing the number of them we can make the energy consuming better. The results of the experiments show that relation (3) decreases the number of cluster heads after each generation. We can see this matter in figure (13).

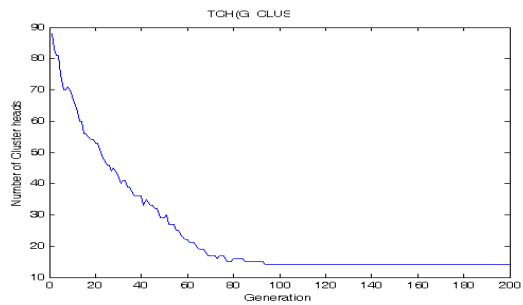

Figure13. The number of cluster heads by the use of relation (3)

Figures (14) and (15) show the relational distance and the amount of energy consuming. They show that our method will decrease these two factors.

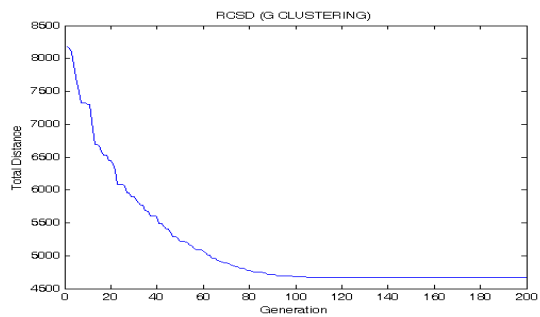

Figure 14. The minimum distance by the use of relation (3)

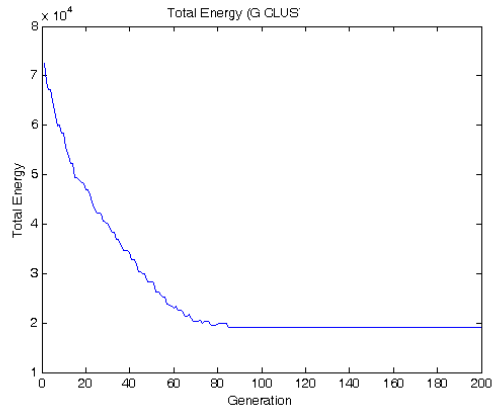

Figure15. The whole consumed energy in a network after each iteration by the help of relation (3)

Figure 16 shows the method of the formation of clusters after doing the genetic algorithm. In this figure, the sink is shown at the middle of the environment with a black square. The regular nodes are shown with blue circles and cluster heads are shown by the yellow circles wich are bigger than the normal nodes.

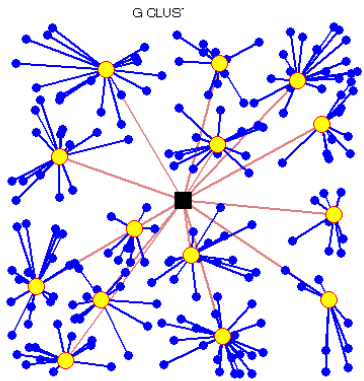

Figure16. the formation of clusters after performing relation (3)

## 5.3. Comparison between suggested method and LEACH algorithm

Figure 17 shows the number of live nodes of a network during 1100 circuits. As we can see in a figure, the sensor nodes in our method stay alive longer than sensor nodes in LEACH.

30

International journal on applications of graph theory in wireless ad hoc networks and sensor networks (GRAPH-HOC) Vol.3, No.1, March 2011

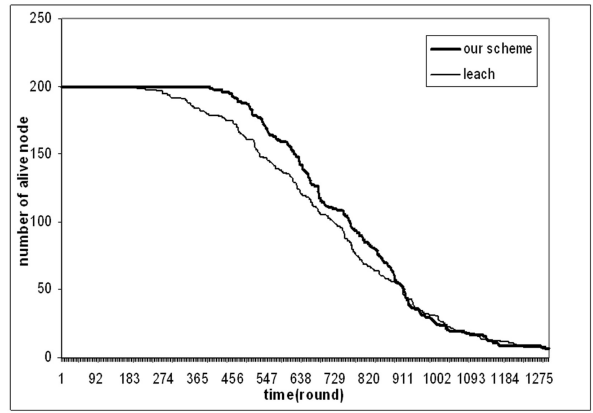

Figure17. The number of live nodes

Figure 18, shows the whole consumed energy in a network during the simulation time. Based on the results of the simulation, more than 18 percent has been economized in consuming energy.

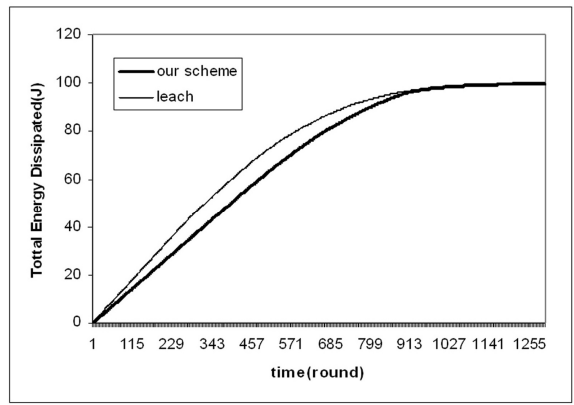

Figure18. The whole consumed energy in a network

## 6. CONCLUSIONS

In this article, one method for clustering the network based on the genetic algorithm was introduced. The basis of this method is the intelligent clustering of network sensors that decrease the connecting distance.

The results of the experiments in this article show that the suggested method is an efficient solution for solving the problem of clusters and their place. It means that the suitable number of cluster heads is determined by the genetic algorithm with a suitable fitness.

Here all the nodes of a network were supposed fixed. But we can change it so that it can be used in networks with movable nodes, too. For determining clusters, we can use the other techniques of learning.